\documentclass[finale]{Wiley}
\usepackage{graphicx}
\usepackage{float}
\usepackage{multicol}
\usepackage{color}
\usepackage{indentfirst}
\usepackage{amsmath} 
\usepackage{siunitx}
\usepackage{ragged2e}
\justifying
\usepackage{bm}
\usepackage{hyperref}
\usepackage{geometry}
\begin{document}
\pagestyle{fancy}
\title{Clay metaBrick-based Motif to Enhance Thermal and Acoustic Insulation}
\maketitle

\author{
Brahim Lemkalli$^{1,2,*}$,
Saad Bensallam$^{3}$,
Muamer Kadic$^{1}$,
Sébastien Guenneau$^{4,5}$,
Hicham Jakjoud$^{6}$,
Abdellah Mir$^{2}$,
Younes Achaoui$^{2}$
}

\vspace{0.5cm}

\begin{affiliations}
$^1$Université Marie et Louis Pasteur, Institut FEMTO-ST, CNRS, 25000 Besançon, France\\
$^2$Laboratory of Optics, Information Processing, Mechanics, Energetics and Electronics, Department of Physics, Moulay Ismail University, B.P. 11201, Zitoune, Meknes, Morocco\\
$^3$Hassania School of Public Works, Casablanca, Morocco\\
$^4$The Blackett Laboratory, Physics Department, Imperial College London, SW7 2AZ, London, UK\\
$^5$UMI 2004 Abraham de Moivre-CNRS, Imperial College London, SW7 2AZ, London, UK\\
$^6$Laboratory of Energy Engineering, Materials and Systems, National Schools of Applied Sciences, Ibn Zohr University, Agadir, Morocco\\
$^{*}$lemkallibrahim@gmail.com
\end{affiliations}


\begin{abstract}
Acoustic metamaterials have gained popularity as promising materials for enhancing noise reduction. Here, we explore the use of metamaterials, based on Helmholtz resonators (HRs), to enhance the performance of standard clay hollow brick. By incorporating HRs in the upper and lower hollows, we transform the standard brick into metaBrick, which is essentially designed based on metamaterial principles. We evaluate the acoustic and thermal performance of walls constructed with clay metaBricks, focusing on sound transmission loss and heat resistance. Both the finite element method and experimental analysis were employed to highlight the performance of metaBricks compared to standard clay bricks. Results show that metaBricks significantly enhance acoustic and thermal insulation, achieving an attenuation of $20$ \si{dB} across a broad frequency range from $500$ to $2500$ \si{Hz} and an $8$\% increase in thermal resistance. However, compressive strength is reduced by $33$\%, though it remains above the standard requirements for building materials. These findings indicate that metaBrick is a promising building material, offering improved sound and thermal insulation.
\end{abstract}

\section{Introduction}

Recently, in response to climate change and noise pollution, building materials with particular characteristics that may be employed to provide efficient heat and sound insulation have become necessary \cite{al2022built, wang2022experimental,anjum2022sustainable}. These materials are crucial for decreasing energy consumption while enhancing comfort and safety in both residential and commercial environments \cite{gao2023comprehensive, sutcu2014thermal, fringuellino1999sound}. Most studies have focused on improving the thermal and acoustic insulation of hollow clay bricks, which are the most commonly used construction materials due to their durability, low cost, and availability \cite{vijayan2021evaluation}. In this regard, several approaches have been explored to increase the insulating properties of clay bricks. A common method is to add insulation materials, such as phase change materials, to the structure of the bricks to improve their thermal resistance \cite{principi2012thermal, mahdaoui2021building, aketouane2018energy, chihab2022thermal, bachir2023numerical, hichem2013experimental, luo2023numerical}. Other research looked at the use of nanocrystalline aluminum sludge to increase the thermal efficiency of perforated clay bricks \cite{santos2015enhancement}. Acoustically, several clay bricks with distinct hollow cores were analyzed \cite{jacqus2011homogenised}. Furthermore, it has been reported that additional layers to the hollow bricks improve their acoustic performance  \cite{souza2020impact}. Another method is to change the structure of the bricks themselves, for example, by modifying the geometry of the air holes \cite{al2017effect}. Besides these approaches, clay brick still has limitations in terms of both thermal and acoustic insulation properties. Therefore, additional advancement in the development of more efficient insulation techniques remains necessary.

To control both phenomena, sound and heat, multi-functional metamaterials that combine these two physics are required. Metamaterials are man-made materials that have been developed to exhibit unique features that are not naturally disposable \cite{Kadic2019, gao2022acoustic, lemkalli2024emergence, lemkalli2023mapping, barhoumi2022improved, huang2023sound,lemkalli2024longitudinal}. These materials have the potential to provide simultaneous heat and sound insulation, which can significantly improve energy efficiency and comfort in buildings and other applications. On the one hand, acoustic metamaterials are a type of metamaterials that are used to manipulate sound waves \cite{cummer2016controlling, sun2020broadband,lemkalli2025controlling, zhang2016three}. In order to improve sound insulation, a variety of resonators with different characteristics are being studied, including phononic crystals with a Bragg-based bandgap \cite{liu2000locally, al2017formation}, free-standing \cite{mei2012dark, yang2015subwavelength, zhang2020light, jang2022lightweight, gu2022laminated}, space coiled \cite{li2016acoustic, cai2014ultrathin, gao2024tunable}, Fabry-Perot \cite{wu2000profiled}, quarter-wavelength \cite{shen2021acoustic}, and Helmholtz resonators\cite{li2016sound,lemkalli2023bi,jimenez2016ultra,jimenez2017rainbow,lemkalli2023lightweight}. By exploiting the special characteristics of acoustic metamaterials, it is feasible to design metamaterials that effectively block, absorb or deflect sound waves. On the other hand, thermal metamaterials have been designed to manipulate heat flow \cite{han2014full, li2021transforming}. These materials have unique thermal properties that enable unprecedented control over heat transfer. By engineering thermal metamaterials, it is possible to create materials that can efficiently trap or release heat, which makes them useful for applications such as thermal insulation \cite{zhang2023diffusion, xu2019metamaterials}. 

Previous studies have also introduced the concepts of metawall and metabrick in the literature. For example, some works have focused on integrating porous materials into metabricks to enhance thermal performance \cite{signanini2019study}. Others have explored metawall unit bricks composed of rubber-metal-concrete composites \cite{mir2018acoustoelastic}. However, the key challenge remains: how can we design a new type of metabrick that provides both thermal and acoustic insulation while relying on commonly used construction materials, as well as those already employed in industrial thermal and sound insulation systems?

In this paper, we suggest a design for a hollow clay brick called a metaBrick, where the word "meta" refers to the concept of metamaterials. The metaBrick is built on Helmholtz resonators, which are designed to reduce the transmission of both sound and heat. Helmholtz resonators are more interesting for creating low-frequency sound insulation \cite{hoppen2023helmholtz}, which is the straightforward reason why we use Helmholtz resonators. To evaluate the acoustic and thermal characteristics of the metaBrick, we combine the finite element method with experimental measurements. The results of the metaBrick are compared to those of the standard clay brick. The first section places a strong focus on the numerical performance analysis of the bricks. The second section is dedicated to empirically evaluating the effectiveness of a wall made of metaBricks, which involves measuring the walls' thermal, acoustic and mechanical properties.

\begin{figure}
    \centering
    \includegraphics[width=0.5\linewidth]{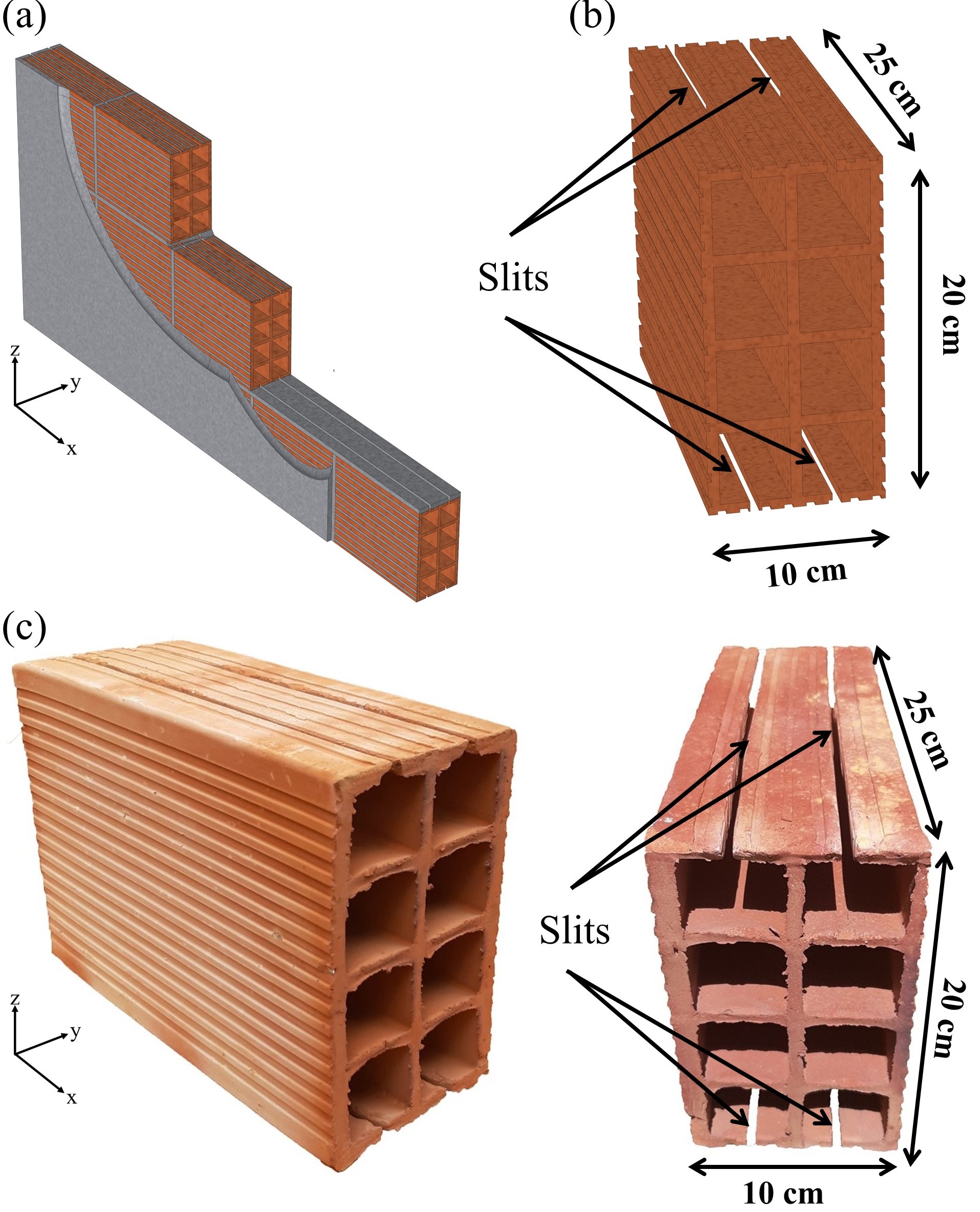}
    \caption{Schematics of the Clay hollow metaBricks. (a) Wall made of clay metaBricks and mortar. (b) Clay hollow metaBrick. (c) Clay hollow metaBrick experimental pictures, with a side view on the left and a top view on the right.}
    \label{Figure 1}
\end{figure}

\section{Characteristics of the metaBrick}

To contribute to the development of the building materials, we have developed the hollow bricks into bricks based on metamaterials called metaBricks, referring to the exotic behavior of metamaterials applied to the specific case of standard hollow bricks. The metaBrick design consists of an array of brick-shaped units with hollow cores, similar to a standard brick, but with the addition of slits at the top and the bottom of the standard hollow brick, as shown in \autoref{Figure 1}. These features create Helmholtz resonators within the metaBrick, which can attenuate sound and reduce heat transfer. The dimensions of the metaBrick used in this study are $10$ \si{cm} in width, $25$ \si{cm} in length, and $20$ \si{cm} in height. 
For the slits on the top and bottom of the metabrick, we conduct a series of simulations (see \autoref{figsup}) to study the effect of slit size on the sound transmission loss. Based on the results, we select slits with a width of $0.5$ \si{cm} and a length of $25$ \si{cm}. The metaBrick-based wall is illustrated in \autoref{Figure 1}(a). The material parameters used in this study are depicted in \autoref{Tab1}.

To comprehensively characterize the acoustic and thermal performances of the metaBrick, we use 2D simulations based on the finite element method. For sound transmission, we use a common method that is based on the excitation of an acoustic plane wave and measures the acoustic transmission of the brick. The response of the brick to the plane wave can provide insights into its ability to absorb, reflect, or transmit sound, which is critical for designing structures that are acoustically comfortable and meet regulatory standards. Similarly, for thermal study, a sinusoidal temperature is often applied to the brick to simulate various thermal conditions. The brick's response to this temperature change is then analyzed to evaluate its thermal performance, including its ability to insulate heat or regulate temperature.

\begin{table}[h]
\caption{\label{Tab1} Building metaBrick mechanical and thermal properties.}
\centering
\setlength{\tabcolsep}{4pt}
\begin{tabular}{l l l l}\hline\hline
Material & Clay& Mortar& Air\\\hline
Young modulus (\si{GPa}) & 17& 25&- \\
Poisson's ratio & 0.2&0.2&-\\
Density (\si{kg/m^3})&2000&2300&1.21 \\
Thermal conductivity (\si{W/(m K)})&0.5&1.8&0.025\\
Specific heat (\si{J/(kg K)})&900&880&1005.3\\
Sound speed (\si{m/s})&-&-&343\\\hline\hline
\end{tabular}
\end{table}

\subsection{Acoustic response}
The numerical model used in the acoustic study is based on solving the acoustic wave propagation equations in both air and solid. Sound waves in air are governed by the Helmholtz equation \ref{eq001} for differential pressure $p$, whereas acoustic waves in the isotropic linear elastic material are governed by the time-harmonic Navier equation (\ref{eq002}):
\begin{equation}\label{eq001}
   \nabla^2 p + \frac{\omega^2}{ C^2}p=0,
\end{equation}
\begin{equation}\label{eq002}
    \frac{E}{2(1+\nu)}(\frac{1}{(1-2\nu)}\nabla (\nabla.\textbf{u})+\nabla^2 \textbf{u})=-\omega^2\rho_s\textbf{u},
\end{equation}
where $\nabla^2$ denotes the Laplacian $\nabla\cdot\nabla$, $p$ is the acoustic pressure, $\textbf{u}$ the displacement field, $C=\sqrt{\frac{1}{\rho_{air}\chi}}$ the speed of sound in air, $\chi$ the air compressibility, $\rho_{air}$ and $\rho_s$ the densities of air and solid, respectively. Besides, $\omega$ is the angular frequency of the acoustic wave, $E$ is the Young's modulus, and $\nu$ is the Poisson's ratio.

A multiphysics model is used to solve the acoustic wave propagation equations in both air and solids for the sound transmission analysis. As part of the multiphysics coupling, we use the following air-solid interface boundary conditions that couple the equations corresponding to waves propagating in both media\cite{yoon2007topology}:

\begin{equation}\label{eq0003}
    \textbf{n}\cdot\frac{1}{\rho_a}\nabla \cdot \textbf{p}=\textbf{n}\cdot\textbf{u}, \;\text{at air-solid interface}
\end{equation}
\begin{equation}\label{eq0004}
    \textbf{n}\cdot \sigma(\textbf{u}) =\textbf{n}\cdot \textbf{p}, \;\text{at solid-air interface}
\end{equation}

where $\sigma$ is the stress tensor for an isotropic linear elastic material, given by $\sigma=\frac{E}{2(1+\nu)}(\frac{1}{(1-2\nu)} \nabla.\textbf{u}+\nabla \textbf{u})$ and $\textbf{n}$ is the outward unit normal to the air or solid.

\begin{figure}[!h]
    \centering
    \includegraphics[width=0.5\linewidth]{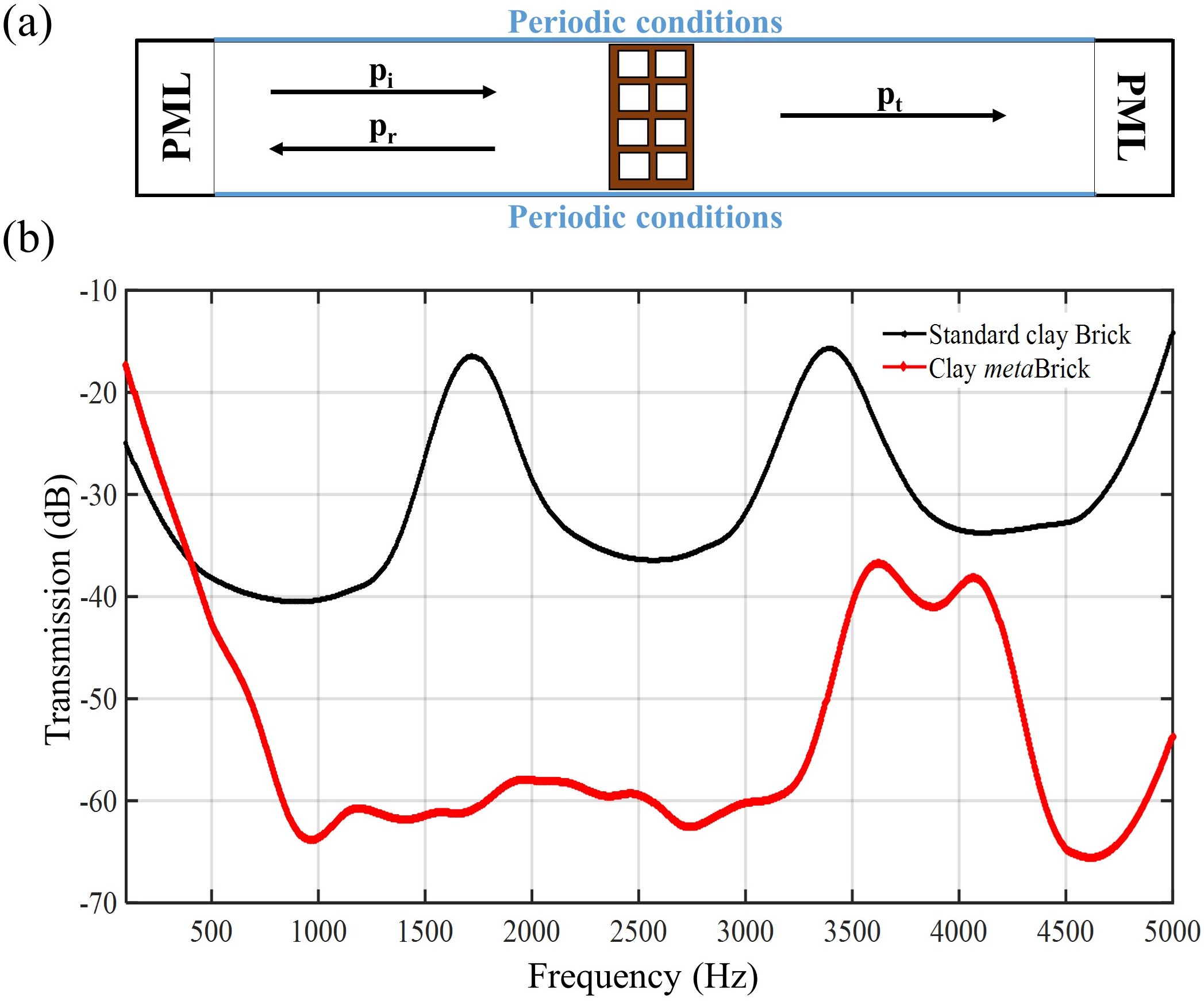}
    \caption{ Sound Transmission. (a) Numerical model. (b) Transmission curves for the two types of bricks: the metaBrick portrayed in the red curve and the standard brick represented by the black curve.}
    \label{Figure 2}
\end{figure}

\begin{figure*}[!ht]
    \centering
    \includegraphics[width=16cm]{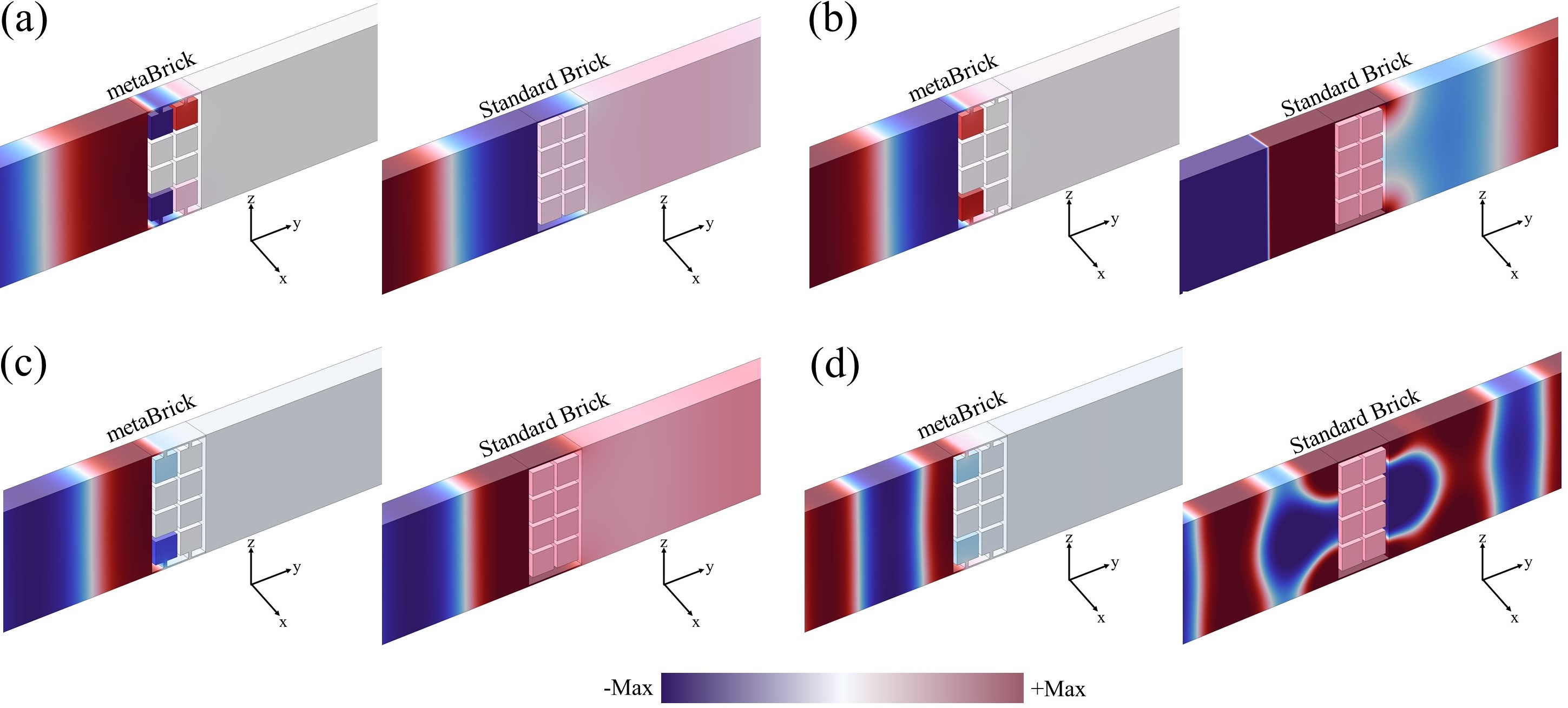}
    \caption{Screenshots of the acoustic response of the standard brick (right) and the metaBrick (left). (a) At a frequency of $500$ \si{Hz}. (b) At a frequency of $671$ \si{Hz}. (c) At a frequency of $793$ \si{Hz}. (d) At a frequency of $1500$ \si{Hz}.}
    \label{Figure 3}
\end{figure*}

In order to simulate the normal incident sound wave, a plane wave radiation boundary condition was utilized, and Perfectly Matched Layers (PMLs) were placed at both extremities of the medium to limit spurious reflections, as depicted in \autoref{Figure 2}(a). Furthermore, the Floquet-Bloch periodic conditions are applied on the top and bottom of the bricks. For the meshing, we take into account a maximum element size of $\lambda_{\text{min}}/10$ and tetrahedral elements. We then calculate the transmission in the frequency range ($20$ \si{Hz} to $5$ \si{kHz}) for each brick, using equation (\ref{eq003}).
\begin{equation}\label{eq003}
    \text{Transmission}=20\times log_{10}(\frac{\iiint|p_t|dV_1}{\iiint |p_i| dV_2}),
\end{equation}
where $V_1$ and $V_2$ are the volumes of the media corresponding to the incident and transmitted waves, respectively, and $p_i$ and $p_t$ represent the incident and transmitted pressures.

\autoref{Figure 2}(b) displays the sound transmission of two types of bricks, the standard brick (black curve) and the metaBrick (red curve), across a wide range of frequencies. It is evident that the standard brick has a higher sound transmission compared to the metaBrick, which is due to the tiny guide between the two successive bricks creating a Fabry-Perot resonator that has a resonance at $n\lambda/2$ (where $\lambda$ is the wavelength). The presence of slits, i.e., Helmholtz resonators, reduces the sound transmission by $20$ \si{dB} across a broad frequency range from $500$ \si{Hz} to $3500$ \si{Hz}, beyond which the limit of diffraction becomes apparent. The superior acoustic insulation properties of the metaBrick are clearly demonstrated through the transmission curve, as it effectively attenuates sound waves more than the classical brick. This finding, which applies without mortar, is of practical significance as it highlights the importance of reducing sound transmission through building materials to create acoustically comfortable structures that comply with regulatory standards.

The response of the two types of bricks to plane wave excitation is further analyzed using snapshots of the acoustic pressure, as shown in \autoref{Figure 3}. For the corresponding displacement fields, see \autoref{figsup2}. The screenshots provide a visual representation of the bricks' reactions to the plane wave at different frequencies and demonstrate that the metaBrick attenuates the sound wave more efficiently than the standard brick. This observation reinforces the conclusion that the metaBrick has superior acoustic insulation properties. The figure particularly highlights the energy trapping inside the Helmholtz resonators in the metabicks at different frequencies, unlike the classical ones, in which cavities play no particular role in the acoustic insulation. In summary, the study provides valuable insights into the acoustic performance of two types of brick and highlights the advantages of using the metaBrick for acoustic insulation applications. 

\subsection{Thermal response}

\begin{figure}
    \centering
    \includegraphics[width=0.5\linewidth]{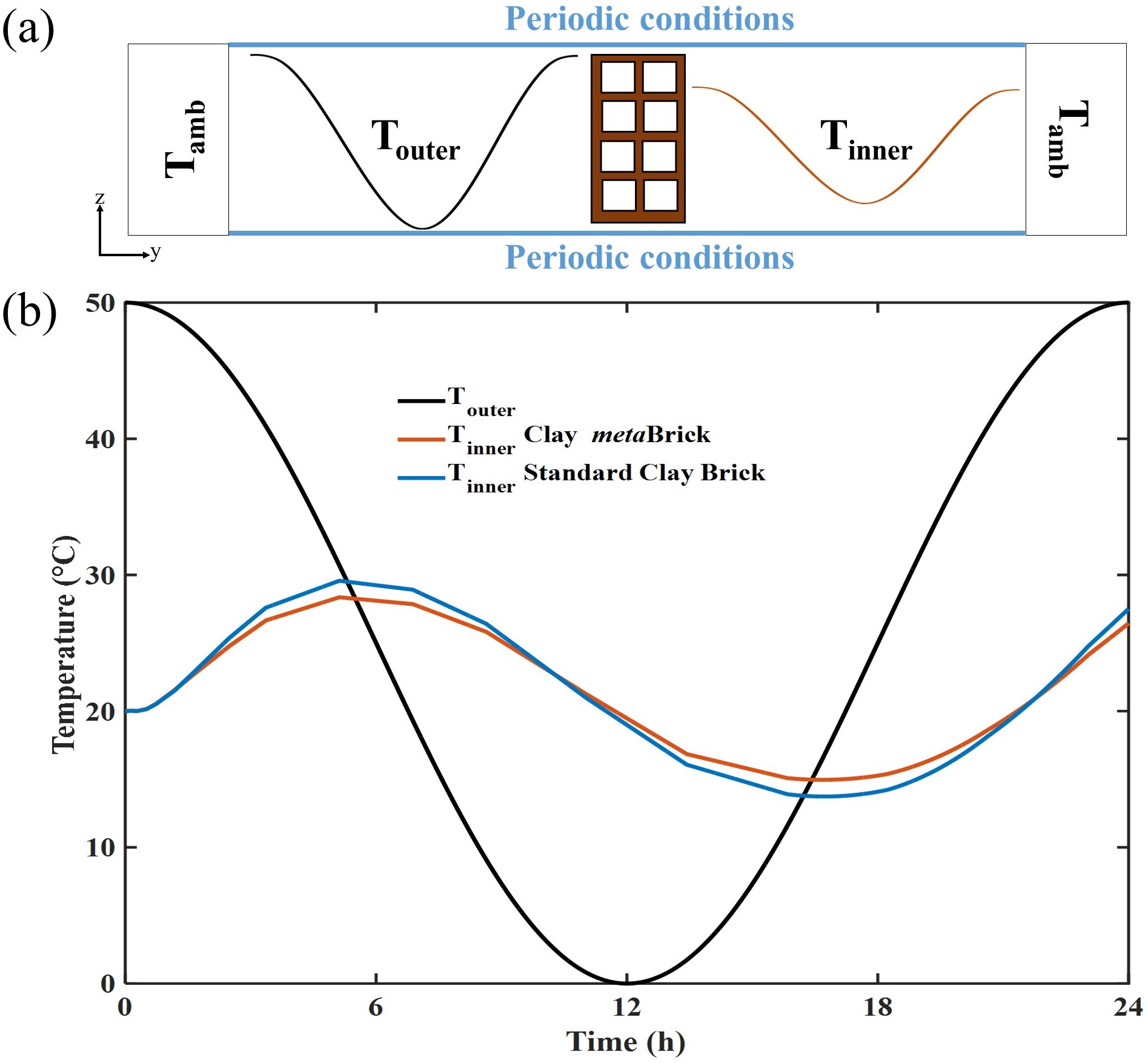}
    \caption{The temperature response. (a) Numerical model. (b) Temperature evolution for the two types of bricks: the standard brick is represented by the blue curve, the metaBrick by the red curve, and the outer temperature by the black curve.}
    \label{Figure 4}
\end{figure}

\begin{figure*}[!ht]
    \centering
    \includegraphics[width=16cm]{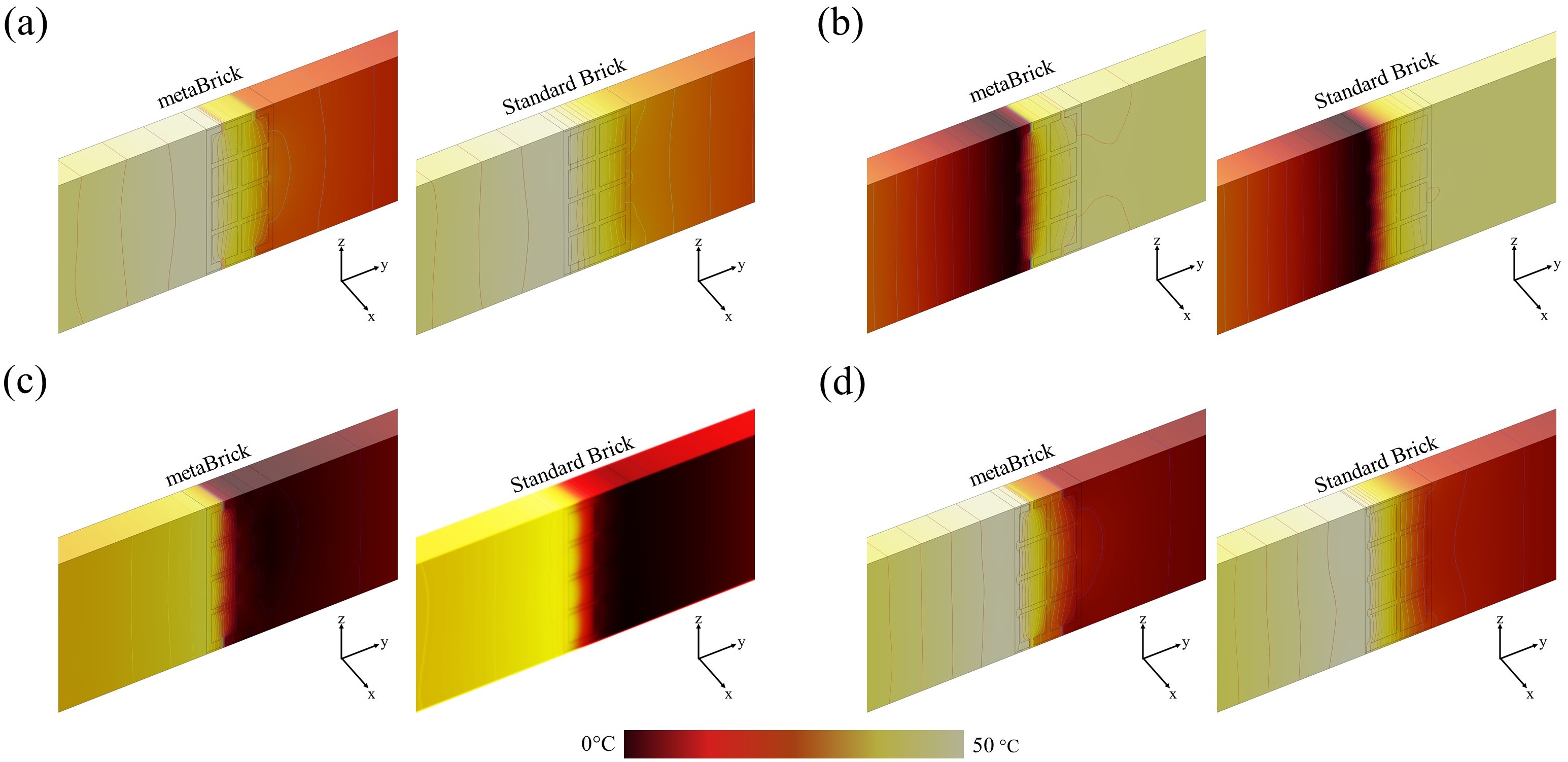}
    \caption{Screenshots of the temperature contour response of the standard brick (right) and the metaBrick (left). (a) After $4$ hours of exposure. (b) After $10$ hours of exposure. (c) After $18$ hours of exposure. (d) After $24$ hours of exposure.}
    \label{Figure 5}
\end{figure*}

For the thermal analysis, the heat transfer in solids and fluids equations was utilized to characterize thermal insulation through the bricks. By solving the transient heat equation (\ref{eq004}), the temperature distribution within the unit cells is calculated. The methodology is based on applying a sinusoidal temperature within a $24$-hour period ranging between $0$ and $50$ \si{\degree C} and determining the inner temperature ($T_{inner}$) of unit cells, as shown in \autoref{Figure 4}(a). To accomplish this, the terminal boundaries of the configuration are exposed to ambient temperature, which is set to $T_{amb}=25$ \si{\degree C}, which means that the inner temperature at the initial time equals to $T_{amb}$. The outer boundary of cells is vulnerable to the outer temperature $T_{outer}(t)$, which describes the temperature variation over a $24$-hour period with the expression $T_{outer}=25 sin(\frac{2\pi}{24}t)+25$ \si{\degree C}. Furthermore, the periodicity conditions are applied in the top and bottom of bricks.
\begin{equation}\label{eq004}
    k_i\Delta T=\rho_i c_i \frac{\partial T}{\partial t},
\end{equation}
where $i$ indicates air or solid, $k_i$, $\rho_i$ and $c_i$ are the thermal conductivity, the density and the specific heat, respectively.

\autoref{Figure 4}(b) reports a comparative analysis of the thermal properties of two types of bricks, namely the standard brick (curve in red) and the metaBrick (curve in blue).  The results demonstrate that the metaBrick exhibits superior thermal insulation properties compared to the standard brick. After $6$ hours of exposure to the sinusoidal outer temperature, the interior temperature of the metaBrick reaches a maximum of $27$ \si{\degree C}, while the standard brick reaches $29$ \si{\degree C}, resulting in a $2$ \si{\degree C} difference in the peak temperature and a $2$ \si{\degree C} difference in the lowest temperature. That means that the metabrick maintains a comfortable internal temperature, making it an excellent choice for thermal insulation over the standard one. The metaBrick, in particular, has lower temperature values than the standard clay brick, suggesting a stronger ability for retaining heat, which can act as an alternative solution to the greenhouse effect.

The superiority of the metaBrick's thermal insulation properties is further supported by the screenshots provided in \autoref{Figure 5}, which depict the temperature distribution across the bricks over time. The screenshots show that the metaBrick retains heat more effectively than the standard brick, thereby reducing the need for additional heating or cooling.

The 2D numerical model provided a useful tool to further investigate the performance of the metaBrick design and its potential to reduce both sound and heat propagation. The simulations performed with the model allowed for the exploration of one scenario where one unit cell represents a brick without mortar and is infinite in the out-of-plane, while in reality the bricks are finite and constructed in a wall with mortar, which would have been difficult or impossible to achieve in the 2D simulations. In the next section, to make the investigation more realistic, the performance of a 3D wall made of bricks and mortar is reported. In order to have a deeper understanding of the acoustic and thermal behaviors of the metaBrick wall, the data obtained from the numerical simulations will be compared to with the data from the experiment.

\section{Characteristics of a metaBrick wall}

Following the 2D numerical simulations that demonstrated the superior acoustic and thermal properties of the metaBrick, our subsequent step involved determining its performance through both experimental and numerical investigations. We constructed two identical walls, one built using the metaBrick and the other using standard brick. The walls were exposed to controlled acoustic and thermal conditions, and we recorded the responses across each wall. By comparing the experimental data obtained from the two walls, we ascertained the performances of the metaBrick wall.

\begin{figure*}[!ht]
    \centering
    \includegraphics[width=12cm]{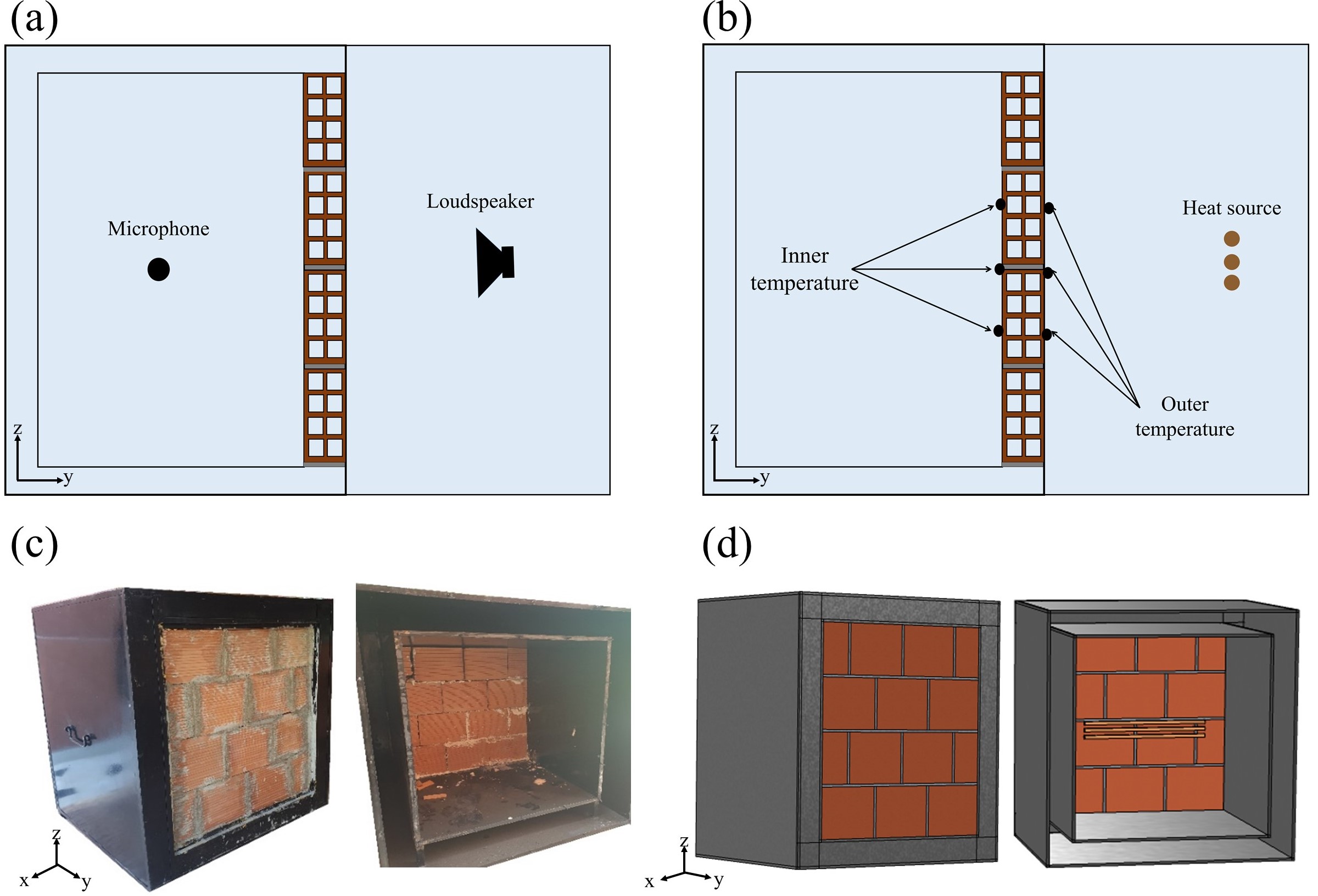}
    \caption{Setups for numerical and experimental analysis. (a) An acoustic setup for measuring sound transmission amplitude in both experimental and numerical studies. (b) The thermal setup to determine the inner temperature in both experimental and numerical studies. (c) The experimental box with a wall. (d) The numerical box with a wall.}
    \label{Figure 7}
\end{figure*}
Furthermore, numerical simulations were carried out to study the behavior of both walls. The results obtained from the simulations were compared with the experimental data to validate the accuracy of the simulation models and gain further insight into the acoustic and thermal properties of the walls. Through the combined approach of experimental and numerical investigations, we were able to obtain a comprehensive understanding of the properties of the metaBrick wall and its advantages over standard brick walls. 
In \autoref{Figure 1}(c), several images of the metaBrick used in our experimental tests are displayed. The images illustrate the top and side views of the metaBrick, side-by-side, allowing for a direct visual comparison of their geometrical attributes. The metaBrick's unique design and composition can be observed, which include a mixture of various materials to enhance its thermal properties. These images provide a visual representation of the experimental setup and the materials used, which can help to understand the experimental results presented further down in this document.

\subsection{Experimental and numerical setups}

The experimental setups used in this study have been mounted to investigate the acoustic and thermal behaviors of the novel metaBrick and compared to a standard brick. The schematics of these setups are illustrated in \autoref{Figure 7}(a) and (b). Thus, these setups were composed of a double steel box, as depicted in \autoref{Figure 7}(c) and (d), in which one side is left open to be filled with the brick-based walls. The walls were constructed using the metaBrick and the standard brick.

For the acoustic study, a loudspeaker was used as the excitation source, and the signals were detected using a microphone that was placed inside the iron box. The theoretical chirp is used to generate the sound, which is at a sufficiently high amplitude in order to get a sound pressure level in the receiving room without the influence of background noise, and the frequency response covers the relevant measurement range, which is at least from $50$ \si{Hz} to $5000$ \si{Hz}. 

The sound pressure level in the box is measured with a calibrated measurement microphone. The microphone was positioned in a fixed location throughout the experiment and numerical test because the sound pressure level in the box varies from one point to another, even in an ideal diffuse sound field. For that, we have conserved the same point for both experimental and numerical tests. The collected data, which represents the pressure in function of time, is processed with the fast Fourier transformation to the amplitude of the sound with the expression $amplitude=20*\log_{10}(p)$. The results of this study are illustrated in \autoref{Figure 8}.

In the thermal study, a heat source was used as the excitation source. Both walls were exposed to an outer temperature of $96$ \si{\degree C} for $600$ minutes in an ambient temperature of $25$ \si{\degree C}, and the temperature fluctuations were measured using an calibrated sensors (infrared thermometers) that were positioned at a fixed distance from the wall surface on both sides, the inner side and the outer side. The thermometers were used to measure the temperature evolution of the wall at different time intervals. The results of this study are illustrated in \autoref{Figure 9}.

The experimental setup allowed for accurate and reliable measurements of the acoustic and thermal behaviors of both the metaBrick and the standard brick. In addition to the experimental analysis, a numerical model was used to further investigate the acoustic and thermal behaviors of the metaBrick design. The numerical model was constructed using the same dimensions and setup as used in the experimental study to ensure consistency and accuracy of the results. Both numerical and experimental setups are based on the same approaches. The objectives of these approaches are to fit the numerical and experimental results and gain a better understanding of the responses of the proposed clay brick. The experimental results were compared with the numerical simulations to validate the accuracy of our modeling approach.

\subsection{Results} 
\begin{figure}[!h]
    \centering
    \includegraphics[width=0.5\linewidth]{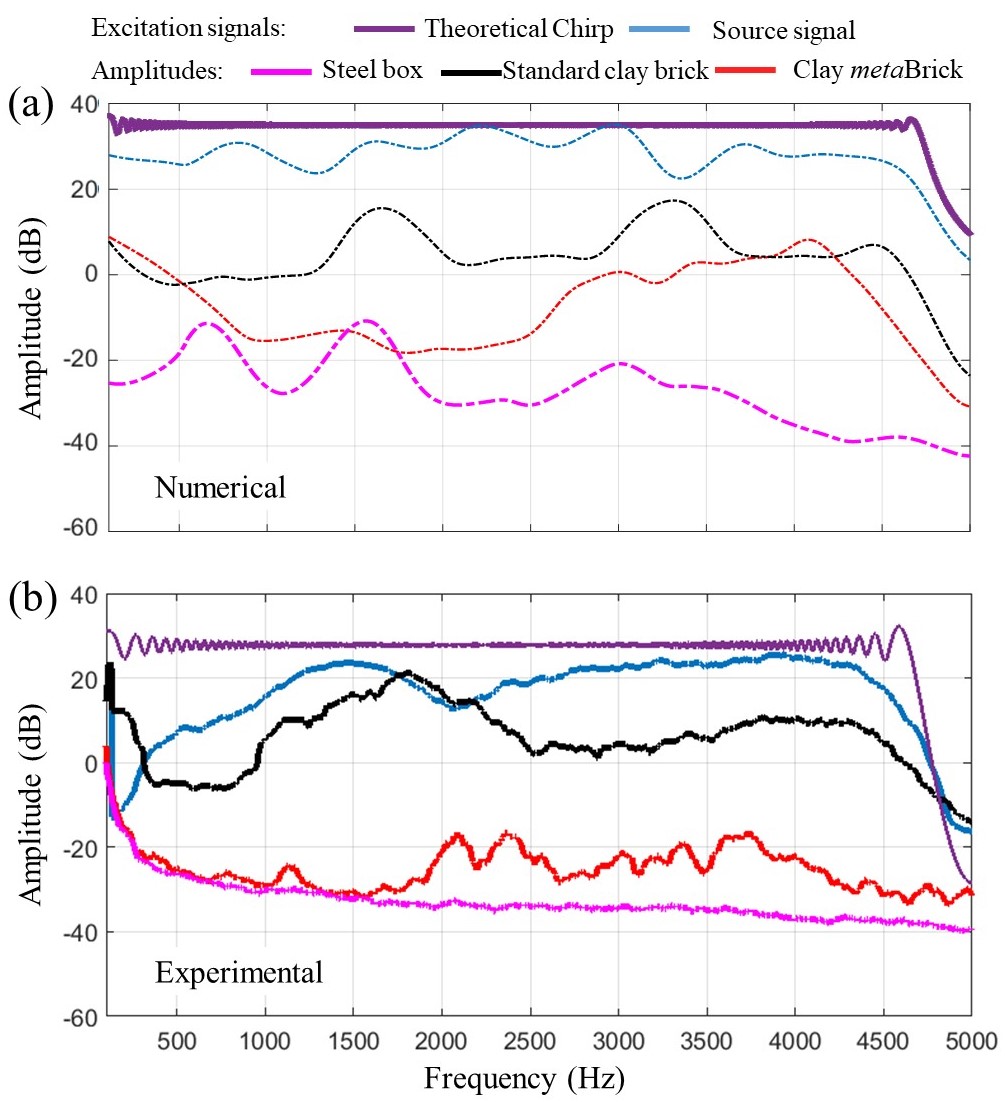}
    \caption{Acoustic responses: the measured excitation signal is represented by the blue curve, the theoretical excitation chirp signal is represented by the violet curve, the steel box sound transmission is represented by the pink curve, the standard brick sound transmission is represented by the black curve, and the metaBrick signal transmission is represented by the red curve. (a) Calculated acoustic responses. (b) Experimental acoustic responses.}
    \label{Figure 8}
\end{figure}

The results obtained from the acoustic analysis provided valuable insights into the effectiveness of the metaBrick design in attenuating sound, as shown in \autoref{Figure 8}. The analysis of the four signals obtained, namely the excitation chirp signal (in violet), the measured excitation signal without walls (in blue), steel box sound transmission (in pink), standard brick sound transmission (in black), and metaBrick signal transmission (in red), enabled a comparison of their respective acoustic behaviors. The numerical and experimental results are presented in \autoref{Figure 8}(a) and (b), respectively. It is clearly seen that the numerical results and the experimental results are in good agreement; in what follows, we interpret both results together.

The steel box transmission signal indicated a very weak transmission amplitude in the frequency range of $100$ to $5000$ \si{Hz}. This observation suggests that the steel box used in the experiment had negligible sound transmission, indicating its suitability for the experiment. This result is validated by the numerical analysis of the steel box transmission.

The standard brick transmission signal, which showed a high transmission curve compared to the steel box, indicated that the sound transmission through the standard brick was high, with the curve closely following the measured excitation signal. It is clear that the amplitude transmission of the standard brick shows a peak in the vicinity of the frequency of $1500$ \si{Hz}, which is produced by the reflections inside the steel box. However, the transmission amplitude of the standard brick shows a transmission loss of $20$ \si{dB} in the band studied in comparison to the measured excitation signal. This finding indicates that the standard brick was not an effective insulator of sound.

In contrast, the transmission signal of the metabrick exhibited a much weaker transmission curve compared to both the measured excitation signal and the transmission of the standard brick. This improvement is primarily attributed to the presence of the slits in the metabrick, which performed exceptionally well in attenuating sound at low frequencies. The transmission loss for the wall constructed with the metabrick reached up to $40$ \si{dB} compared to the measured excitation signal and $20$ \si{dB} compared to the transmission loss of the standard brick. However, both the numerical and experimental results show some discrepancies, which may be due to differences in the material properties used in the simulations and the real-world scenario. For the numerical models, it is important to note that we assumed all materials to be linear isotropic elastic, which does not reflect real-world conditions. In practice, additional factors such as porosity, viscosity, and other material characteristics can influence the results.

\begin{figure}[!h]
    \centering
    \includegraphics[width=0.5\linewidth]{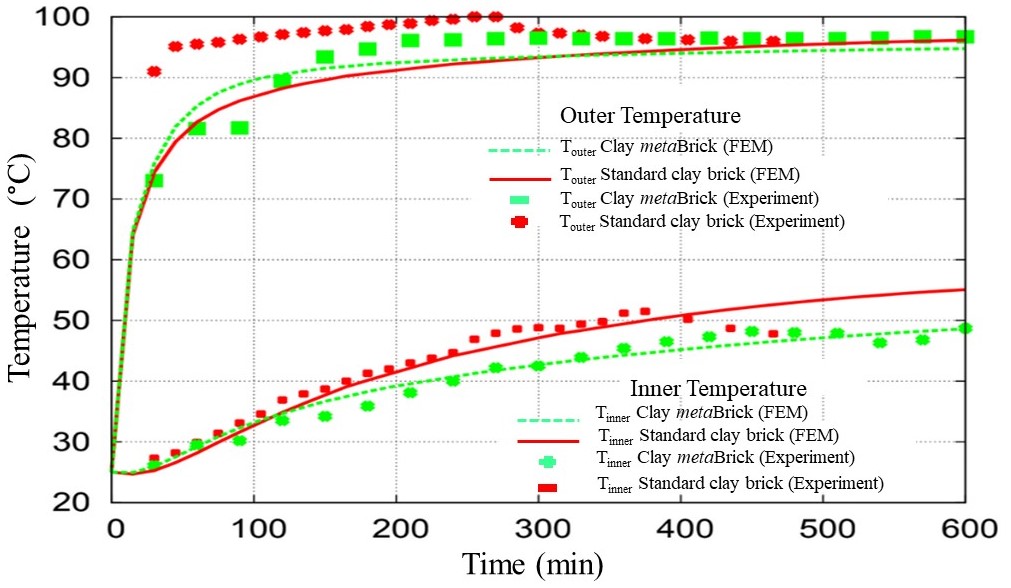}
    \caption{Numerical and experimental inner and outer temperatures change as a function of exposure time.}
    \label{Figure 9}
\end{figure}

To elaborate further, the thermal study involved measuring the temperature variation inside and outside the iron box over a period of 600 minutes while exposed to a temperature of $90$ \si{\degree C}. Inner and outer temperature evolution is measured for both numerical and experimental studies. The results of this study are presented in \autoref{Figure 9}. It is clearly seen that the numerical and the experimental results are in good agreement.

The inner temperature variation showed that the metaBrick wall was highly effective in thermal insulation, as evidenced by the lower variation of the inner temperature compared to the standard brick. After 600 min of exposure to the outer temperature, as one can see, the standard brick wall exhibited a temperature variation of up to $55$ \si{\degree C} ($38\%$ of reduction), while the metaBrick wall exhibited an inner temperature of up to $48.5$ \si{\degree C} ($46\%$ of reduction). Thus, make a difference of $6.5$ \si{\degree C}, which is equivalent to $8\%$ as a rate of thermal reduction. Furthermore, the outer temperature variation was also significantly lower than that of the standard brick, indicating its effectiveness in thermal insulation. 

\section{Discussion}

To compare the acoustic and thermal properties of the metaBrick with a standard brick, we used finite element simulations and experimental measurements. We found that the presence of the slits in the metaBrick design created a barrier, which effectively reduced the sound transmission and heat diffusion. The presence of the Helmholtz resonators in the metaBrick design allowed for the dissipation of sound and heat energy, reducing the amplitude of the transmitted waves. This is in contrast to a standard brick, which does not have the ability to attenuate sound or heat in this way. The acoustic insulation becomes higher with a sound transmission of up to $20$ \si{dB} in the frequency range of $200$ \si{Hz} to $3.5$ \si{kHz}. The same occurs for the thermal insulation, which became more efficient with a rate of $8\%$ compared with the standard brick. Moreover, bricks are usually subjected to compression loads. As a result, understanding the compressive strength of bricks is of the utmost significance for establishing their suitability for construction. 

We conducted experimental compression tests on both the metabrick and the standard clay brick. Four samples of each brick type were tested to determine their compressive strength, expressed in \si{MPa}. The results are summarized in \autoref{Tab2}.


\begin{table}[h]
\caption{\label{Tab2} Experimental compressive strength of bricks.}
\centering
\begin{tabular}{l l l}\hline\hline
Type of brick & Sample & Compressive strength (\si{MPa}) \\\hline
& 1 & 19.5 \\
& 2 & 17 \\
Standard clay brick & 3 & 21 \\
& 4 & 20\\ 
& 5 & 19\\\hline
& 1 & 11\\
& 2 & 14.5 \\
Clay metaBrick & 3 & 12.5 \\
& 4 & 13 \\
& 5 & 12 \\\hline\hline
\end{tabular}
\end{table}

The standard brick has an average compressive strength of around $19$ \si{MPa}, whereas the metaBrick has an average compressive strength of about $12$ \si{MPa}. Although metaBrick has lost strength, it still possesses a relatively high compressive strength that is more than enough for the building sector ($7.5$ \si{MPa}) \cite{gencel2015characteristics}. The reason for the strength loss is due to the existence of slits in metaBrick, which causes the compressive strength to decrease by approximately $7$ \si{MPa}, which is equivalent to $33\%$ when compared with standard brick.

This metaBrick results in more comfortable and environmentally friendly buildings, reducing energy consumption and noise transmission in the construction industry, and is easy to manufacture by just adding some steel beams to the existing molds in the industries. The presence of the Helmholtz resonators in the metaBrick design allows for a significant reduction in sound transmission and thermal conductivity, making it a promising candidate for use in building materials applications. These findings can be useful for researchers, designers, and engineers who seek to develop more sustainable and environmentally friendly building materials that offer better acoustic and thermal insulation properties.

The concept of metamaterials is a relatively new path in material design, and our study shows that it has immense potential for revolutionizing the construction industry. The use of metamaterials can pave the way for the development of more sustainable and energy-efficient building materials, leading to a reduction in energy consumption and a more comfortable building environment. Our study's findings have significant practical implications and can contribute to the development of innovative solutions for the construction sector.

\section{Conclusions}
In conclusion, the present study has investigated the acoustic and thermal behaviors of a metamaterial design named metaBrick. Numerical simulations and experimental measurements have demonstrated that the presence of slits in alveoli in the metaBrick structure creates Helmholtz resonators, which can effectively attenuate both sound and heat. The findings of this study indicate that metaBrick has a strong potential for improving insulation performance in building materials with a compressive strength that is high enough for the construction sector. The results also indicate that further research is warranted to explore the use of metamaterials in multi-functional applications to enhance building performance and energy efficiency. Overall, this study has demonstrated the importance of developing new and innovative building materials to meet the increasing demand for sustainable and energy-efficient buildings. Metamaterials, such as the metaBrick design, have shown potential for making significant contributions to this effort.

\section*{Acknowledgments}
We would like to thank the Moroccan Ministry of Higher Education, Scientific Research and Innovation and the OCP Foundation who funded this work through the APRD research program. We acknowledge the support of the ANR PNanoBot project (contract "ANR-21-CE33-0015"). 

\section*{Appendix A: Sound transmission as a function of slit width variation.}
\begin{figure}[h!]
    \centering
    \includegraphics[width=0.5\linewidth]{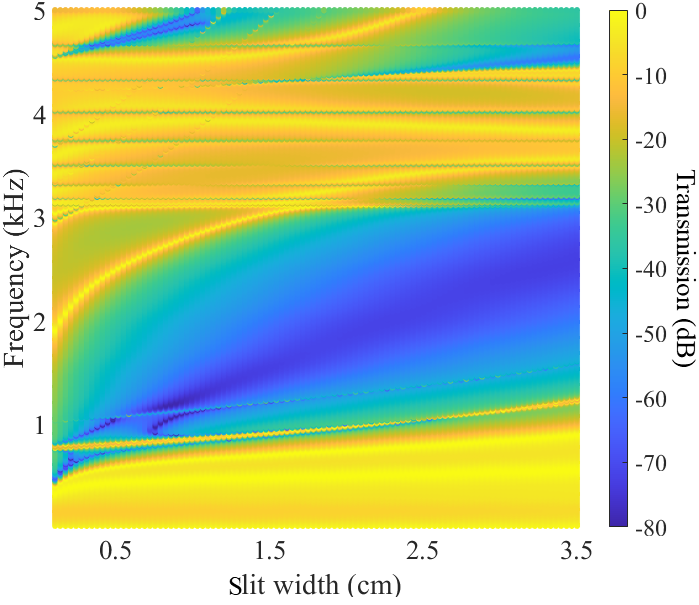}
    \caption{Sound transmission as a function of slit width variation.}
    \label{figsup}
\end{figure}
\newpage
\section*{Appendix B: Displacement distribution in the metabrick.}
\begin{figure}[h!]
    \centering
    \includegraphics[width=1\linewidth]{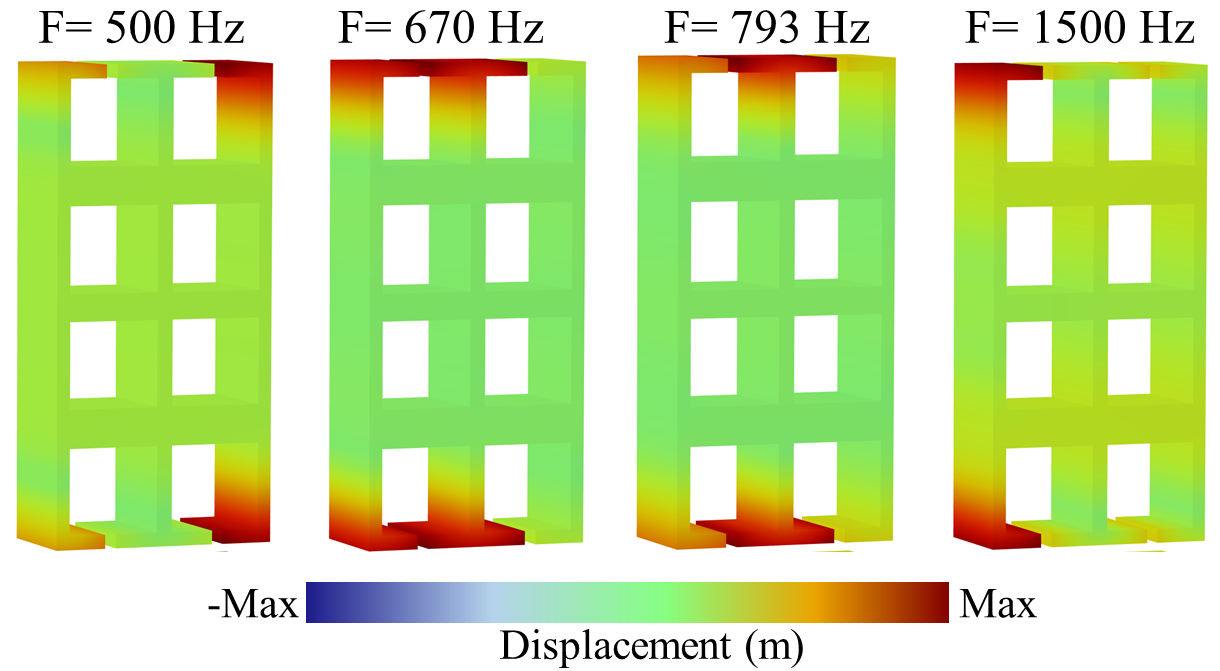}
    \caption{{Screenshots of the displacement modulus of the metaBrick. (a) At a frequency of 500 Hz. (b) At a frequency of 671 Hz. (c) At a frequency of 793 Hz. (d) At a frequency of 1500 Hz.}}
    \label{figsup2}
\end{figure}

\bibliographystyle{MSP}
\bibliography{mybibfile}

\end{document}